\documentstyle[12pt,epsfig]{article}
\def\nn{\nonumber}
\def\ee{\end{equation}}
\def\be{\begin{equation}}
\def\eea{\end{eqnarray}}
\def\bea{\begin{eqnarray}}
\def\eeas{\end{eqnarray*}}
\def\beas{\begin{eqnarray*}}
 
\begin{document}
\begin{titlepage}
\begin{center}
{\large\bf The pairing Hamiltonian for one pair of nucleons bound
in a potential well }
\vspace*{1cm}

{M.B.Barbaro$^{1,2}$, L.Fortunato$^1$, A.Molinari$^{1,2}$ 
and M.R.Quaglia$^{1,2}$}

\vspace*{1cm}
{\it 
$^1$ Dipartimento di Fisica Teorica dell'Universit\`a di Torino, 
via P.Giuria 1, I--10125, Torino, Italy\\
$^2$ INFN Sezione di Torino}

\vspace*{1.5cm}

{\bf ABSTRACT}\\

\begin{quotation}
{\it The problem of one pair of identical nucleons sitting in ${\cal N}$
single particle levels of a potential 
well and interacting through the pairing force 
is treated introducing, in the Hamiltonian 
formalism, even Grassmann variables. The eigenvectors are analytically 
expressed solely in terms of these with coefficients fixed by the eigenvalues 
and the single particle energies. For these a specific model is needed: 
in the case of the harmonic oscillator well, for any strength of the pairing 
interaction, an accurate expression is derived
for both the collective eigenvalue and for those 
trapped in between the single particle levels.
Notably the latter are labelled through an index upon which they depend 
parabolically.}
\end{quotation}
\end{center}


\end{titlepage}

We have recently obtained, in the framework of the Grassmann algebra, 
the analytic expressions for the eigenvalues and the eigenvectors 
of $n$ pairs of like-nucleons interacting through the pairing Hamiltonian
and sitting in one single-particle level \cite{Bar00}.

When the single-particle levels, with angular momentum 
$j_1, j_2, \cdots j_{\cal N}$ and energies
$e_1, e_2, \cdots e_{\cal N}$ (all the $j$'s being assumed to be different),
are ${\cal N}$ usually the problem is dealt with numerically. 
In this case the Hamiltonian of the system reads

\bea
H= \sum_{\nu=1}^{\cal N} e_\nu \sum_{m_\nu=-j_\nu}^{j_\nu} 
\lambda^*_{j_\nu m_\nu} \lambda_{j_\nu m_\nu} - 
G \sum_{\mu,\nu=1}^{\cal N} 
\sum_{m_\mu=1/2}^{j_\mu} \lambda^*_{j_\mu m_\mu} \lambda^*_{j_\mu {\bar
 m_\mu}}
\sum_{m_\nu=1/2}^{j_\nu} \lambda_{j_\nu {\bar m_\nu}} \lambda_{j_\nu  m_\nu}
\ ,\nonumber
\\
\label{Ham}
\eea
where $\lambda_{jm}$ and $\lambda^*_{jm}$ are the odd (anticommuting, 
nilpotent) Grassmann variables 
(associated to the nucleons annihilation and creation, respectively) and

\be
\lambda_{j {\bar m}} \equiv (-1)^{j-m} \lambda_{j -m} \ .
\ee

Introducing even (commuting, nilpotent) Grassmann variables to describe a 
pair of fermions with vanishing third component of the total angular momentum
($M$=0), namely \cite{Bar97}

\be
\varphi_{j m} \equiv (-1)^{j-m} \lambda_{j -m} \lambda_{j m} \ ,
\label{phi}
\ee
(\ref{Ham}) becomes

\be
H= \sum_{\nu=1}^{\cal N} e_\nu \sum_{m_\nu=-j_\nu}^{j_\nu} 
\lambda^*_{j_\nu m_\nu} \lambda_{j_\nu m_\nu} - 
G \sum_{\mu,\nu=1}^{{\cal N}} 
\sum_{m_\mu=1/2}^{j_\mu}\sum_{m_\nu=1/2}^{j_\nu}
 \varphi^*_{j_\mu m_\mu}  \varphi_{j_\nu  m_\nu}
\label{hamiltonian} \ .
\ee

In this letter we provide the eigenvectors of one pair of 
fermions interacting 
through the Hamiltonian (\ref{hamiltonian}) in the presence of ${\cal N}$ 
single particle levels in terms of (\ref{phi}) with coefficients expressed 
through the eigenvalues and the single particle energies. 
When the latter are those of an harmonic oscillator, an accurate, almost 
analytic expression is given for {\em all} the eigenvalues as well.

We start by counting the total 
number $N_s$ of states available to the system: it is given by

\be
N_s={2\Omega\choose 2} = N_s^{(1)} + N_s^{(2)}= 
\frac{1}{2}\sum_{\mu\ne\nu=1}^{\cal N} 
2\Omega_\mu \cdot 2\Omega_\nu + 
\sum_{\nu=1}^{\cal N} {2\Omega_\nu\choose 2}
\label{Ns}
\ee
where
\be
\Omega=\sum_{\nu=1}^{\cal N}\Omega_\nu= \sum_{\nu=1}^{\cal N} (j_\nu+1/2)\ .
\ee
In the above the states with the two fermions sitting on 
two different levels are $N_s^{(1)}$, those 
with the two fermions placed on the same level are $N_s^{(2)}$.
Since we are interested in the physics where the 
pairing force is active, we consider of the $N_s^{(2)}$ states only the ones
having the two fermions in time reversal orbits: these are 
$ \sum_{\nu=1}^{\cal N} \Omega_\nu$.

Notwithstanding the presence of  
both the $\lambda$'s and the $\varphi$'s in (\ref{hamiltonian}), 
we search for eigenstates in the form

\be
\psi = \sum_{\nu=1}^{{\cal N}} \sum_{m_\nu=1/2}^{j_\nu} 
\beta_{j_\nu m_\nu} \varphi^*_{j_\nu m_\nu} \ .
\label{psi}
\ee
Using the Grassmann algebra rules, the coefficients 
$\beta$ are then found to obey the system of $\Omega$ equations 

\bea
\left ({\cal E}-2 \varepsilon_\nu\right) \beta_{j_\nu m_\nu} + 
\sum_{{\mu}=1}^{{\cal N}} \sum_{m_\mu=1/2}^{j_\mu} 
\beta_{j_\mu m_\mu} =0 \ ,
\label{eqS}
\eea
where $1\leq\nu\leq {\cal N}$, $1/2\leq m_\nu\leq j_\nu$.

We cast the above system in the $\Omega\times\Omega$ matrix form

\bea
{\bf M}\vec\beta =
\left(
\begin{array}{cccccc}
B_{11} & B_{12} & \cdot & \cdot & \cdot & B_{1{\cal N}} \\
B_{21} & B_{22} & \cdot & \cdot & \cdot & B_{2{\cal N}} \\
\cdot  & \cdot  & \cdot & \cdot & \cdot & \cdot  \\
\cdot  & \cdot  & \cdot & \cdot & \cdot & \cdot  \\
\cdot  & \cdot  & \cdot & \cdot & \cdot & \cdot  \\
B_{{\cal N}1} & B_{{\cal N}2} & \cdot & \cdot & \cdot & B_{{\cal N}{\cal N}} 
\end{array}
\right)
\left(
\begin{array}{c}
\beta_{j_1 1/2}\\
\beta_{j_1 3/2}\\
\cdot\\
\cdot\\
\cdot\\
\beta_{j_{\cal N} j_{\cal N}}
\end{array}
\right)=
 0\ ,
\eea
where the elements of 
the blocks $B_{\mu\nu}$, of dimensions $\Omega_\nu\times\Omega_\mu$, are
$[B_{\mu\nu}]_{\alpha \beta}=
\left({\cal E}-2 \varepsilon_\nu\right)\delta_{\mu\nu}\delta_{\alpha\beta}+1$.

In the above all the energies are expressed in
units of $G$, i.e.
\be
\varepsilon_\nu = e_\nu/G \ \mbox{and} \ \ {\cal E}= E/G\ .
\label{calH}
\ee

The matrix ${\bf M}$ is easily diagonalized, the associated determinant being
\be
\det {\bf M} =
\left[\prod_{\rho=1}^{\cal N}
({\cal E}-2\varepsilon_\rho)^{\Omega_\rho-1} \right]
\left[\prod_{\mu=1}^{\cal N}({\cal E}-2 \varepsilon_\mu)
+\sum_{\nu=1}^{\cal N} \Omega_\nu \prod_{\mu\neq\nu=1}^{\cal N} ({\cal E}-2 
\varepsilon_\mu)
\right]
\label{eigen}
\ee
and the corresponding secular equation admits $2{\cal N}$ 
distinct solutions. 
Clearly ${\cal N}$ of these are 
\be
{\cal E}_\rho= 2 \varepsilon_\rho \ ,\ \mbox{with degeneracy}\ 
d_\rho=\Omega_\rho-1\ ,\ \ 1\le \rho \le {\cal N}\ ,
\label{free}
\ee
the remaining  ${\cal N}$ 
(non-degenerate) fulfilling instead the equation \cite{Rowe}

\be
1+f({\cal E})=0\ \ \ 
\mbox{with}\ \ \ 
f({\cal E})=\sum_{\nu=1}^{\cal N} 
\frac{\Omega_\nu}{{\cal E}-2 \varepsilon_\nu}  \ .
\label{v0}
\ee
The eigenvalues (\ref{free}) clearly 
correspond to states insensitive to the pairing interaction, being associated 
to a pair with non vanishing angular momentum\footnote{Obviously 
if $\Omega_\nu=1$, then 
$d_\nu=0$: hence the free solution (\ref{free}) is absent. 
Indeed a pair of fermions on the level $j_\nu=1/2$ can only couple 
to angular momentum $J=0$, hence feeling the pairing interaction.}.
The solutions of (\ref{v0}) correspond instead to states 
(describing a pair coupled to an angular momentum $J=0$) 
affected by the pairing force. 

Now, beyond the trivial solutions (\ref{free}),
also the eigenvalues fulfilling (\ref{v0}) 
and the associated eigenstates can be analytically obtained (at least, 
as we shall see, for an
harmonic oscillator potential well) in a basis, of dimensions lower than 
(\ref{phi}), which generalizes the one we introduced 
in ref.\cite{Bar00}.

To show how this occurs we organize, as in ref.\cite{Bar00}, 
the states associated with each of the
${\cal N}$ single particle levels into two sets, 
embodying $\Omega_\nu-1$ and 1 levels, respectively
(of course it must be $\Omega_\nu\ne 1$).
Correspondingly we define the following new basis of $2{\cal N}$ 
normalized states 
$\Phi^{(a)}_{\nu}$,
with $1\le\nu\le {\cal N}$ and $a=0,1$:

\be
\left\{
\begin{array}{cclc}
\Phi^{(0)}_\nu & = & \frac{1}{\sqrt{d_\nu}}
\sum_{m_\nu=1/2}^{j_\nu-1}\varphi_{j_\nu,m_\nu}
\ , & \ \ \ \ \ \nu=1,\cdots {\cal N}
\\
\Phi^{(1)}_\nu &=& \varphi_{j_\nu,j_\nu} &
\end{array}
\ .\right. 
\label{Phi}
\ee

In this basis the rectangular blocks $B_{\mu\nu}$ become
\bea
B_{\mu\nu}=\left(
\begin{array}{cc}
B^{(0)}_{\mu\nu} & {\bf a}^T_{\nu} \\
{\bf a}_{\mu} & B^{(1)}_{\mu\nu}
\end{array}
\right) \ ,
\label{Bmunu}
\eea
where the row vector ${\bf a}_{\mu}$, of dimension $(\Omega_\mu-1)$, is
filled with ones
and $B_{\mu\nu}^{(0)}$ and $B_{\mu\nu}^{(1)}$ are two matrices of dimensions
$(\Omega_\nu-1)\times(\Omega_\mu-1)$  and
$1\times 1$, respectively.

Then the $2{\cal N}\times 2{\cal N}$ matrix ${\bf{\cal M}}$, 
representing the operator ${\cal H}= {\cal E}-H/G$ 
in the basis (\ref{Phi}), is obtained 
by replacing each $B_{\mu \nu}$ with a $2 \times 2 $ matrix with elements 
given by the sum of the elements of each block in (\ref{Bmunu})
(divided by the corresponding normalization factors).  
One thus gets 
\be
{\bf{\cal M}}=\left(
\begin{array}{cccccc}
{\cal E}-2 \varepsilon_1+d_1 & \sqrt{d_1} & 
 \cdot & \cdot & \sqrt{d_1 d_{\cal N}} & \sqrt{d_1}\\
\sqrt{d_1}  & {\cal E}-2 \varepsilon_1+1 &
\cdot & \cdot &
\sqrt{d_{\cal N}} & 1 \\
\cdot  & \cdot  & \cdot & \cdot & \cdot & \cdot  \\
\cdot  & \cdot  & \cdot & \cdot & \cdot & \cdot  \\
\sqrt{d_{\cal N} d_1}  & \sqrt{d_{\cal N}}  & \cdot & \cdot & 
{\cal E}-2 \varepsilon_{\cal N}+d_{\cal N}& 
\sqrt{d_{\cal N}}  \\
\sqrt{d_1} &1& 
 \cdot &\cdot & \sqrt{d_{\cal N}} &
{\cal E}-2 \varepsilon_{\cal N}+1
\end{array}
\right) \ ,
\label{calM}
\ee
whose determinant reads

\bea
\det {\bf{\cal M}}&=&
\det \left(
\begin{array}{cccccc}
1+\frac{{\cal E}-2 \varepsilon_1}{d_1}& 1 & \cdot & \cdot & 1 & 1\\
1  & 1+{\cal E}-2 \varepsilon_1  & \cdot & \cdot & 1 & 1 \\
\cdot  & \cdot  & \cdot & \cdot & \cdot & \cdot  \\
\cdot  & \cdot  & \cdot & \cdot & \cdot & \cdot  \\
1  & 1  & \cdot & \cdot &1+\frac{{\cal E}-2 \varepsilon_{\cal N}}{d_{\cal N}} 
& 1  \\
1 &1& \cdot &\cdot & 1 &  1+{\cal E}-2 \varepsilon_{\cal N}
\end{array}
\right)\prod_{\nu=1}^{\cal N} d_\nu 
\nonumber\\
&=&
\left[\prod_{\rho=1}^{\cal N}
({\cal E}-2\varepsilon_\rho) \right]\left[\prod_{\mu=1}^{\cal N}({\cal E}-2 
\varepsilon_\mu)
+\sum_{\nu=1}^{\cal N} \Omega_\nu \prod_{\mu\neq\nu=1}^{\cal N} ({\cal E}-2 
\varepsilon_\mu)
\right]\ .
\label{detM}
\eea

From (\ref{detM}) both the ${\cal N}$ unperturbed solutions 
${\cal E}_\rho = 2\varepsilon_\rho$ and
the ${\cal N}$ ones affected by the pairing interaction follow. 

The latter are also obtained in the normalized basis 
(introduced long ago by Richardson \cite{Ric64} with a different technique)

\be
\widetilde\Phi_\nu = \sqrt{1-\frac{1}{\Omega_\nu}} \left[ 
\Phi^{(0)}_\nu+ \frac{1}{\sqrt{d_{\nu}}}\Phi^{(1)}_\nu\right]\ ,
\ \ \ \ \ \ \ \ \ \nu=1,\cdots {\cal N}\ ,
\label{Phitilde}
\ee
which linearly combines the building blocks of (\ref{Phi}).
Note that the index of nilpotency of the commuting variables 
$\widetilde\Phi_\nu$ is $\Omega_\nu$: hence the (\ref{Phitilde}) are ``more 
bosonic'' than the (\ref{phi}). Indeed they are often referred to as 
$s$-quasibosons ($J=0$) and lead to the following ${\cal N}$ 
dimensional representation of ${\cal H}$ 
\footnote{An alternative option uses for  
the basis the unnormalized vectors 
$\Phi'_{\nu}= \Omega_{\nu}\widetilde{\Phi}_{\nu}$. 
In this case the ${\cal N}$-dimensional matrix representing ${\cal H}$ 
is filled with
ones everywhere, except in the principal diagonal.}

\be
{\bf{\widetilde {\cal M}}}=\left(
\begin{array}{cccccc}
{\cal E}-2 \varepsilon_1+\Omega_1 & \sqrt{\Omega_1\Omega_2} & 
 \cdot & \cdot & \cdot & \sqrt{\Omega_1\Omega_{\cal N}}\\
\sqrt{\Omega_1\Omega_2}  & {\cal E}-2 \varepsilon_2+\Omega_2 &
\cdot & \cdot &\cdot & \sqrt{\Omega_2\Omega_{\cal N}} \\
\cdot  & \cdot  & \cdot & \cdot & \cdot & \cdot  \\
\cdot  & \cdot  & \cdot & \cdot & \cdot & \cdot  \\
\cdot  & \cdot  & \cdot & \cdot & \cdot & \cdot  \\
\sqrt{\Omega_1 \Omega_{\cal N}}  & \sqrt{\Omega_2\Omega_{\cal N}}  
& \cdot & \cdot & \cdot & 
{\cal E}-2 \varepsilon_{\cal N}+\Omega_{\cal N}
\end{array}
\right) \ .
\ee

It is worth noticing that Richardson's basis, being of lower dimension, 
only yields the so-called seniority $v=0$ solutions. 
In contrast, our basis (\ref{Phi}) provides the whole spectrum (any seniority)
of the pairing hamiltonian eigenvalues: in the present case, of course, 
the $v=2$ 
solutions are trivial, but when several pairs are present the $v\ne 0$ 
solutions are important. 

In general the roots of Eq.(\ref{v0})
cannot be given analytically for ${\cal N}>4$. Only the lowest 
{\em collective} eigenvalue, when $G$ is large, 
turns out to be (in dimensional units)
\bea
E_c \stackrel{G\to\infty}{\sim} - G\sum_{\nu=1}^{\cal N} \Omega_\nu 
= -G\Omega\ 
\label{coll}
\eea
for any potential well. 

The other ${\cal N}-1$ solutions, as is well-known, get instead ``trapped''
between the unperturbed energies

\bea
2 e_{\nu-1} < E_\nu < 2  e_{\nu}
\ \ \ \ \ \ \ \ \ \nu=2,\cdots {\cal N} \ 
\eea
and in the limit $G\to\infty$ are given by the zeros of 
$f(E/G)$ and depend upon the ${\cal N}$ single 
particle energies $e_\nu$ and their degeneracies $\Omega_\nu$.

Remarkably, when one degeneracy, say $\Omega_\nu$, becomes very large, 
then the  ``trapped'' eigenvalues, for {\it any} value of $G$, 
coincide with the free ones, i.e.

\be
\lim_{\Omega_\nu\to\infty} E_\mu =
 \left\{
\begin{array}{l}
 2e_{\mu-1}
\ \ \ \ \ \ \ \mbox{for}\ \ 2\leq\mu\leq\nu
\nn\\
 2e_{\mu}
\ \ \ \ \ \ \ \mbox{for}\ \ \nu<\mu\leq {\cal N}\ ,
\end{array}
\right.
\ee
whereas the lowest collective energy tends to $-\infty$.

Concerning the eigenfunctions, those (normalized) corresponding to the 
eigenvalues ${\cal E}_\nu=2\varepsilon_\nu$ $(\nu= 1,\cdots {\cal N})$ 
(seniority $v=2$ states) in the basis (\ref{Phi}) are

\be
\psi_{\nu,v=2}(\Phi^*)=\sqrt{1-\frac{1}{\Omega_\nu}}\left\{
\left[\Phi_\nu^{(1)}\right]^*-\frac{1}{\sqrt{d_\nu}} 
\left[\Phi_\nu^{(0)}\right]^*\right\}
\label{psiv2}
\ee
and describe a free pair sitting in the level $j_\nu$.

The $v=0$ eigenfunctions are more conveniently expressed in the 
basis (\ref{Phitilde}). Here they read 

\be
\psi_{v=0}(\widetilde\Phi^*) = \sum_{\nu=1}^{\cal N} 
\sqrt{\Omega_\nu} w_\nu \widetilde\Phi^*_\nu\ ,
\ee
the coefficients $w_\nu$ fulfilling the system of equations

\be
\left( {\cal E} - 2 \varepsilon_\nu \right) w_\nu + 
\sum_{\mu=1}^{\cal N} \Omega_\mu w_\mu = 0 \ .
\label{system}
\ee
The above is easily solved and yields the noticeable formula
\be
w_{\nu}= \frac{{\cal E}-2\epsilon_{\cal N}}{{\cal E}-2\epsilon_{\nu}}
w_{{\cal N}}\ .
\label{w}
\ee
Since (\ref{w}) entails for large ${\cal E}$ 
\be
w_1=w_2=\cdots=w_{\cal N} \ ,
\ee
the collective eigenstate in the $G \rightarrow \infty$ limit 
corresponds to a coherent superposition of all the $s$-quasibosons reading 
\be
\psi_{v=0}(E_c) \stackrel{G\to\infty}
{\sim} \sum_{\nu=1}^{\cal N} 
\sqrt{\Omega_\nu}\widetilde\Phi^*_\nu 
= \sum_{\nu=1}^{\cal N}
\sum_{m_\nu=1/2}^{j_\nu} \varphi^*_{j_\nu m_\nu}\ ,
\label{psiv0}
\ee
which is completely symmetric in the exchange of any pair of $\varphi$,
thus exhibiting the same symmetry of the pairing Hamiltonian.

Also from (\ref{w}) one sees that, 
in the limit $G \rightarrow 0$ where
${\cal E}_{\nu}\simeq 2\epsilon_{\nu}$, only one component of the basis, 
i.e. the $\nu$-th one, survives in the wavefunction of the ``trapped'' 
states. The same occurs 
when one degeneracy, say $\Omega_{\nu}$, becomes very large:
as for the eigenvalues,
the eigenvectors then coincide with the $\mu$-th component of the basis 
for $\mu > \nu$ and with the $(\mu-1)$-th component for $\mu \le \nu$.

We now choose the  
harmonic oscillator potential for the single particle energies 

\be
e_N = \left(N+\frac{3}{2}\right)\hbar\omega \ \ \ \ \ \ 
N=0,\cdots,\infty
\label{eho}\  
\ee
and for the degeneracies of the states available to a pair
\be
\Omega_N=\frac{1}{2}(N+1)(N+2)
\label{deg} \ .
\ee

In this instance the secular equation (\ref{v0}), 
when the lowest ${\cal N}$ levels are considered,
becomes
\be
\sum_{N=0}^{{\cal N}-1}\frac{(N+1)(N+2)}{2 N+3-\widetilde{E}}=
\frac{1}{\widetilde{G}}\ 
\label{eqho}
\ee
where $\widetilde{G}=G/2\hbar\omega $
and the energies are measured in units of $\hbar \omega$ 
$(2 {\widetilde{e}}_N=2N+3)$.

\begin{figure}[ht]
\centerline{
\psfig{figure=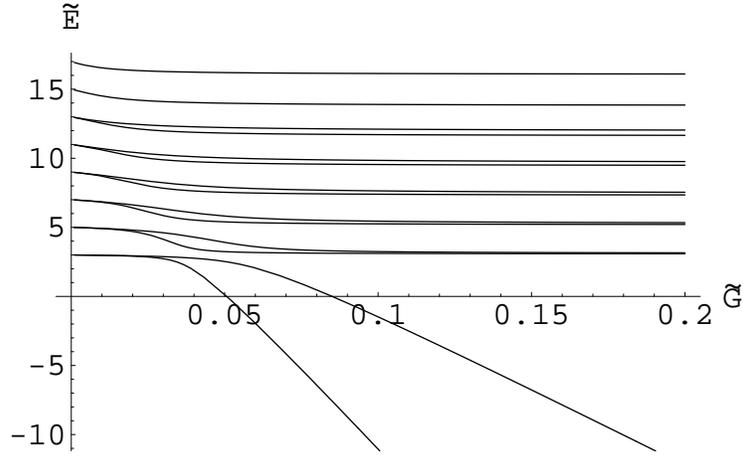,width=10cm,clip=}}
\caption[ ]{The figure shows the solutions $\widetilde E$ of Eq.(\ref{eqho}),
for ${\cal N}$=6 (upper curves) and 8 (lower curves), 
as functions of $\widetilde{G}$.
One can see that with the harmonic oscillator well each trapped solution 
$\widetilde{E}_N$ for $\widetilde{G}>0.1$ tends approximatively to the single 
particle energy $2\widetilde{e}_N$.}
\end{figure}

In Fig.1 the numerical solutions of (\ref{eqho}) are displayed 
for ${\cal N}$=6 and 8 versus $\widetilde{G}$. 
Remarkably, the dependence upon $\widetilde{G}$ is lost 
for $\widetilde{G}\ge 0.1$. Furthermore in this regime the number of trapped
solutions is obviously fixed by ${\cal N}$, but their dependence
upon ${\cal N}$ is quite mild.

We now conjecture the ${\cal N}$ trapped solutions of 
(\ref{eqho}), which we label with the index $\bar N$, to 
{\it depend parabolically upon the latter}, namely

\be
\widetilde{E}_{\bar N}=a {\bar N}^2+b {\bar N}+c 
\ \ \ \ \ \ \ \ \ \ \ \ \ \ {\bar N}=0,\cdots {\cal N}-2
\label{parab}
\ee
(the collective solution $\widetilde{E}_{c}$ will be separately treated). 

To fix the coefficients $a$, $b$ and $c$, we recast (\ref{eqho}) in the 
polynomial form 

\be
\widetilde{E}^{\cal N}+a_1 \widetilde{E}^{{\cal N}-1}+
a_2 \widetilde{E}^{{\cal N}-2}+
\cdots a_{{\cal N}-1} \widetilde{E}+a_{\cal N}=0 
\label{eqk}
\ee
and compute the first three coefficients. They turn out to be 
\bea
a_1 &=& \frac{1}{3} {\cal N}({\cal N}+2) [\widetilde G({\cal N}+1)-3]
\\
a_2 &=&  \frac{1}{6} {\cal N}({\cal N}-1) 
[-\widetilde G ({\cal N}+1)({\cal N}+2)(2{\cal N}+3)+
3{\cal N}^2+11 {\cal N}+11]
\\
a_3 &=&  \frac{1}{90} {\cal N}({\cal N}-1)({\cal N}^2-4) 
[\widetilde G({\cal N}+1)(15 {\cal N}^2+40{\cal N}+27)-
15({\cal N}^2+3 {\cal N}+3)]\ .
\nonumber
\\
\eea

Then the first three Viete equations, namely

\bea
\sum_{\bar N=0}^{{\cal N}-2} \widetilde{E}_{\bar N} &=& -a_1-\widetilde{E}_c
\label{Vieta1}
\\
\sum_{\bar N=0}^{{\cal N}-2} \widetilde{E}^2_{\bar N} &=& 
a_1^2-2 a_2-\widetilde{E}_c^2
\label{Vieta2}
\\
\sum_{\bar N=0}^{{\cal N}-2} \widetilde{E}^3_{\bar N} &=& 
-3 a_3-a_1(a_1^2-3 a_2)- 
\widetilde{E}_c^3 \ ,
\label{Vieta3}
\eea
yield a non-linear system in the unknowns $a$, $b$ and $c$, if
$\widetilde{E}_c$ is known. 
This system can be solved by expressing, via eq.(\ref{Vieta1}), 
$c$ as a function of $a$ and $b$ 
\bea
c(a,b)&=&-\frac{1}{{\cal N}-1}\left\{\widetilde{E}_c+\frac{b}{2} ({\cal N}-1)
({\cal N}-2)+\frac{a}{6} ({\cal N}-1)({\cal N}-2)(2{\cal N}-3) \right.
\nonumber\\
&+&
\left.\frac{1}{3} {\cal N}({\cal N}+2)
\left[{\widetilde G}({\cal N}+1)-3\right]\right\}\ .
\label{eqc}
\eea
In turn (\ref{eqc}), inserted into (\ref{Vieta2}), yields $b$ as 
a function of $a$. One finds
\be
b(a) = -\frac{ 15 a({\cal N}^4 -6 {\cal N}^3+13 {\cal N}^2-12 {\cal N}+4)
\pm\sqrt{\Delta}}{15 ({\cal N}-1)^2 ({\cal N}-2)}
\label{eqb0}
\ee
with
\bea
\Delta &=&  -15 ({\cal N}-1)^2 ({\cal N}-2) \left\{ a^2 ({\cal N}-1)^2
({\cal N}-1)({\cal N}-2)({\cal N}-3)
\right.\\
 &+& 20 \left[9 \widetilde{E}_c^2 + 6 \widetilde{E}_c
({\cal N}+2)(\widetilde{G}{\cal N}+\widetilde{G}-3)-3({\cal N}^3-4 {\cal N}^2
-13 {\cal N}-11)
\right.\nonumber\\
 &-& \left. \left. \widetilde{G}^2 {\cal N}({\cal N}-2)({\cal N}+1)^2
({\cal N}+2)^2 + 3 \widetilde{G}({\cal N}+1)({\cal N}+2)({\cal N}^2-
4 {\cal N}-3)\right]\right\}\ .
\nonumber
\label{eqb}
\eea

\begin{figure}[ht]
\centerline{
\psfig{figure=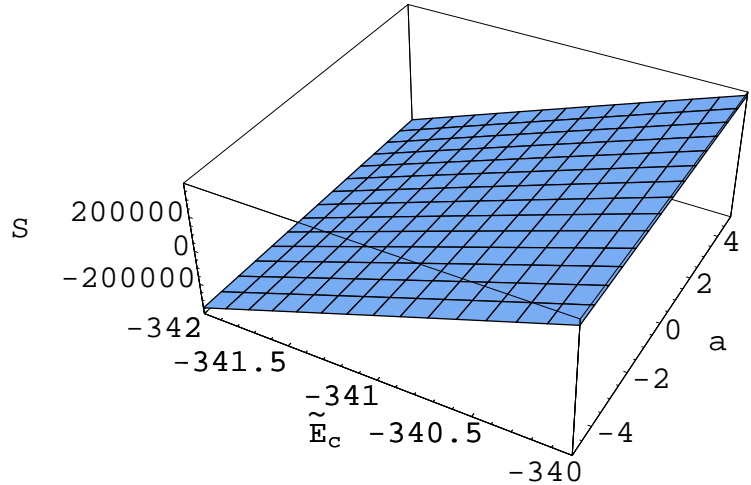,width=10cm,height=8cm,clip=}}
\caption[ ]{The figure shows the surface $S(a,\widetilde{E}_c)$
for $-342<\widetilde{E}_c<-340$ and  $-5<a<5$. Here the number of levels
is ${\cal N}=5$ and the pairing coupling constant is $\widetilde{G}=5$.}
\end{figure}
 
Finally, from (\ref{Vieta3}), an equation for $a$ follows, not reported here, 
being quite cumbersome. 
While we have analytically solved (\ref{Vieta1}) and (\ref{Vieta2}) 
(they are of first and second degree in $c$ and $b$, respectively), 
the non linear equation
(\ref{Vieta3}) for $a$ can only be solved numerically. 
Note that in (\ref{eqb0})
the plus sign in front of the square root should be taken: however 
both options, when inserted in (\ref{Vieta3}), lead to the same
equation for $a$.

It should be stressed that, because of the high degree of non-linearity of 
the latter, its solutions, {\it when} $\widetilde G$ {\it is large},
turn out to be extremely sensitive to the collective energy $\widetilde{E}_c$.
To illustrate this issue we display in Fig.2,  
for ${\cal N}=5$ and $\widetilde{G}=5$, the surface (see (\ref{Vieta3}))
\be
S(a,\widetilde{E}_c)=
\sum_{\bar N=0}^{{\cal N}-2} \widetilde{E}^3_{\bar N} +3 a_3+a_1(a_1^2-3 a_2)+ 
\widetilde{E}_c^3 \ 
\ee
whose zeros clearly give the values of the parameter $a$ entering 
into (\ref{parab}). 
It appears from the figure that even a tiny variation of 
$\widetilde{E}_c$ induces a gigantic variation in $S$, implying that 
$\widetilde{E}_c$ should be fixed with extreme precision in order to obtain 
the correct results for $a$, $b$ and $c$.

Now a very good expression for the collective energy, obtained
by expanding Eq.(\ref{eqho}) in the parameter $(2N+3)/\widetilde{E}$ and 
retaining the leading order, reads

\be
\widetilde{E}_c^{(0)} = -\frac{\widetilde{G}}{3} {\cal N}({\cal N}+1)
({\cal N}+2) +\frac{3}{2} ({\cal N}+1) \ .
\label{Ecappr}
\ee

\begin{table}[hbt]
\begin{center}
\begin{tabular}{c|c|c|c|c|c|c}
 & \multicolumn{3}{|c|}{$\widetilde G=1$}& 
\multicolumn{3}{|c}{$\widetilde G=5$}\\ \hline
${\cal N}$ & $\widetilde{E}_c^{(e)}$  &  $\widetilde{E}_c^{(0)}$  
& $\widetilde{E}_c^{(1)}$ & $\widetilde{E}_c^{(e)}$  &  
$\widetilde{E}_c^{(0)}$  & $\widetilde{E}_c^{(1)}$        \\
\hline
2          &   -3.6055513   &  -3.5   &    -3.6041143 
           & -35.5192213        & -35.5  &   -35.5192121  \\ 
3          &   -14.096330   &  -14.0  &    -14.095861  
           &   -94.0182425        & -94.   &   -94.0182392  \\
4          &   -32.582181    &  -32.5  &   -32.582012 
           &   -192.515882        & -192.5 &   -192.515881  \\
5          &   -61.070599    &  -61.0  &   -61.070528 
           &   -341.013793        & -341.  &   -341.013793  \\
6          &   -101.56156    &  -101.5 &   -101.56152 
           &   -549.512104        & -549.5 &   -549.512104  \\
7          &   -156.05444   &  -156.0 &    -156.05442
           &   -828.010748        & -828.  &   -828.010748  \\
8          &   -226.54874    &  -226.5 &   -226.54873 
           &   -1186.50965        & -1186.5&   -1186.50965  \\
\end{tabular}
\caption{Comparison between the exact $(e)$ and the approximate $(0)$ 
(eq.(\ref{Ecappr})) and $(1)$ (eq.(\ref{Ecappr1})) collective energies  
for some values of ${\cal N}$ and $\widetilde{G}$=1 and 5.}
\end{center}
\end{table}
Table 1 demonstrates the validity of (\ref{Ecappr}). Yet,
its level of precision
is not sufficient to allow a reliable determination, through (\ref{Vieta3}), 
of $a$, because of the dramatic non-linearity displayed in Fig.2.
We are thus forced to improve upon (\ref{Ecappr}). To this purpose we  
insert into (\ref{eqho}) the following expression for the collective energy
\be
\widetilde{E}_c = \widetilde{E}_c^{(0)} + \delta 
\ee
and again expand in the very small parameter $\delta/M(N)$ where
\be
M(N)= 2N+3 -\widetilde{E}_c^{(0)}\ .
\ee
We thus obtain, 
{\it in lieu} of (\ref{Ecappr}), the expression 
\bea
\widetilde{E}_c^{(1)}&=& \widetilde{E}_c^{(0)} +
\left[\frac{1}{\widetilde G}- 
\sum_{N=0}^{{\cal N}-1}\frac{(N+1)(N+2)}{2N+3-\widetilde{E}_c^{(0)}}\right]
\left[\sum_{N=0}^{{\cal N}-1}\frac{(N+1)(N+2)}{
{\left(2N+3-\widetilde{E}_c^{(0)}\right)}^2}
\right]^{-1}
\nn\\
&=& \widetilde{E}_c^{(0)} -
\frac{{\cal L}({\cal N}, \widetilde E_c^{(0)}) + 2{\cal N}
\left( \widetilde{E}_c^{(0)}+ {\cal N} +2\right)-8/\widetilde{G}}
{ 2 {\cal N} + 
\frac{\partial {\cal L}({\cal N}, \widetilde E_c^{(0)})}
{\partial \widetilde{E}_c^{(0)}} }\ ,
\label{Ecappr1}
\eea
where 
\be
{\cal L}({\cal N}, \widetilde E_c^{(0)})\equiv
\left[ {\left(\widetilde{E}_c^{(0)}\right)}^2 -1\right]\left[ 
\Psi\left( {\cal N}+ \frac{3-\widetilde{E}_c^{(0)}}{2}\right)-
\Psi\left( \frac{3-\widetilde{E}_c^{(0)}}{2}\right)\right] 
\ee
($\Psi$ being the Digamma function). Eq.(\ref{Ecappr1})
provides the exact collective energy, 
for all practical purposes, as shown in the fourth and last column of Table 1.

However, for weaker $\widetilde G$, (\ref{Ecappr1}) becomes less 
accurate, because the expansion parameter $(2N+3)/\widetilde E$ is no longer 
small. But this occurrence is of no consequence for the trapped solutions 
(up to about $\widetilde G=1$), since a weaker $\widetilde G$ also means 
a less severe non linearity. If, however, a more accurate value for the 
collective energy is wished, it can be found through the recursion relation 
\be
\widetilde{E}_c^{(k+1)}=\widetilde{E}_c^{(k)} +
\left[\frac{1}{\widetilde G}- 
\sum_{N=0}^{{\cal N}-1}\frac{(N+1)(N+2)}{2N+3-\widetilde{E}_c^{(k)}}\right]
\left[\sum_{N=0}^{{\cal N}-1}\frac{(N+1)(N+2)}{
{\left(2N+3-\widetilde{E}_c^{(k)}\right)}^2}
\right]^{-1}\ ,
\label{Erecursion}
\ee  
the index $k$ labelling the order of the iteration. 
We have numerically checked that the iterative expansion (\ref{Erecursion}) 
converges to the exact solution for $\widetilde G> 0.2$. 

However, the physical interesting domain for the pairing problem occurs 
for $\widetilde G=0.05 - 0.1$, since \cite{Fesh} 
\be
G\simeq\left\{\begin{array}{ll}
27/A \ \mbox{MeV} & \mbox{for protons}\\
22/A \  \mbox{MeV} & \mbox{for neutrons}
\end{array}
\right.
\ee
$A$ being the nuclear mass number, and 
$\hbar \omega$ should correspond to the average 
distance between the single particle levels inside the last occupied shell,
namely $\approx 0.69$~MeV in Pb and $\approx 0.74$~MeV in Sn.

Interestingly an accurate expression for the collective energy can 
also be given 
in the regime $0\le \widetilde G \le 0.2$. Indeed here 
of ${\widetilde E}_c(\widetilde G, {\cal N})$ we know the value 
for $\widetilde{G}=0$ (namely 3) and  
where it vanishes, namely for 
\be
\widetilde{G}_0=\left[\sum_{N=0}^{{\cal N}-1}\frac{(N+1)(N+2)}
{\left(2N+3\right)}\right]^{-1}\ . 
\ee
\begin{table}[hbt]
\begin{center}
\begin{tabular}{c|c|c|c|c|c|c}
& \multicolumn{6}{|c}{$\widetilde G=0.05$}\\ \hline
${\cal N}$ & $\widetilde{E}_c^{(0)}$  &  $\widetilde{E}_c^{(1)}$  
& $\widetilde{E}_c^{(2)}$  & $\widetilde{E}_c^{(3)}$ & $\widetilde{E}_c^{(4)}$
& $\widetilde{E}_c^{(e)}$       \\
\hline
2    & 2.87617 & 2.88394 & 2.88349 & 2.88348 & 2.88348  & 2.88348  \\ 
3    & 2.81180 & 2.87628 & 2.86196 & 2.86014 & 2.86012  & 2.86012  \\
4    & 2.70198 & 2.89185 & 2.84845 & 2.82471 & 2.82056  & 2.82046  \\
5    & 2.53580 & 2.86975 & 2.80333 & 2.75484 & 2.74104  & 2.74028  \\
6    & 2.24594 & 2.63571 & 2.54951 & 2.52958 & 2.52886  & 2.52886  \\
7    & 1.61587 & 1.84837 & 1.83108 & 1.83094 & 1.83094  & 1.83094  \\
8    & .134714 & .136135 & .136135 & .136135 & .136135  & .136135 \\
\hline
& \multicolumn{6}{|c}{$\widetilde G=0.1$}\\ \hline
${\cal N}$ & $\widetilde{E}_c^{(0)}$  &  $\widetilde{E}_c^{(1)}$  
& $\widetilde{E}_c^{(2)}$  & $\widetilde{E}_c^{(3)}$ & $\widetilde{E}_c^{(4)}$
& $\widetilde{E}_c^{(e)}$       \\
\hline
2          &  2.70757   &  2.72972 & 2.72822 & 2.72822 & 2.72822 &  2.72822 \\ 
3          &  2.45586   &  2.61626 & 2.58336 & 2.58335 & 2.58335 &  2.58335  \\
4          &  1.96617   &  2.21918 & 2.18238 & 2.18238 & 2.18238 &  2.18238 \\
5          &  0.919942  &  1.00125 & 1.00000 & 1.00000 & 1.00000 &  1.00000 \\
6          & -1.66966   & -1.47625 &-1.48025 &-1.48025 &-1.48025 & -1.48025  \\
7          & -8.37559   & -4.88888 &-5.44566 &-5.44568 &-5.44568 & -5.44568  \\
8          & -24.4931   & -3.26859 &-10.7011 &-11.0491 &-11.0546 & -11.0546  \\
\end{tabular}
\caption{Comparison between the exact $\widetilde E_c^{(e)}$ and 
the approximate $\widetilde E_c^{(k)}$ 
[eq.(\ref{Ecubic}) and eq.(\ref{Erecursion})] collective energies  
for some values of ${\cal N}$ and $\widetilde{G}$=0.05 and 0.1}
\end{center}
\end{table}
Moreover, the value of its derivative is $-2$ in $\widetilde G=0$ and 
\be
\left.\frac{\partial \widetilde E_c}{\partial \widetilde G}
\right |_{{\widetilde G}_0} =
-\left[\sum_{N=0}^{{\cal N}-1}\frac{(N+1)(N+2)}{2N+3}\right]^2
\left[\sum_{N=0}^{{\cal N}-1}\frac{(N+1)(N+2)}{
{\left(2N+3\right)}^2}\right]^{-1}
\ee     
in $ \widetilde G= {\widetilde G}_0$.

These four constraints are fulfilled by the cubic 
\be
\widetilde E_c^{(0)}(\widetilde G)= 3-2\widetilde G -
\left\{9+\left[
\left.\frac{\partial \widetilde E_c}{\partial \widetilde G}
\right |_{ {\widetilde G}_0}
-4\right]\widetilde G_0\right\}
\frac{\widetilde G^2}{\widetilde G_0^2} + 
\left\{6+\left[
\left.\frac{\partial \widetilde E_c}{\partial \widetilde G}
\right |_{{\widetilde G}_0}
-2\right]\widetilde G_0 \right\}
\frac{\widetilde G^3}{\widetilde G_0^3}\ , 
\label{Ecubic}
\ee
which thus provides an excellent starting point for the evaluation of the 
collective energy. Proceeding indeed as done in the large $\widetilde G$ 
domain, a perturbative expansion in a parameter $\delta$ can be set up, 
leading again to formula (\ref{Ecappr1}), but with the input 
$\widetilde E_c^{(k)}$ now given by the {\it k}-th iteration of 
(\ref{Ecubic}). 
How remarkably accurate the results we obtain are, is displayed in Table~2. 

\begin{table}
\begin{center}
\begin{tabular}{c|c|c|c|c|c|c|c|c}
 & \multicolumn{2}{|c|}{$\widetilde G=0.05$}& 
\multicolumn{2}{|c}{$\widetilde G=0.1$}
 & \multicolumn{2}{|c|}{$\widetilde G=1$}& 
\multicolumn{2}{|c}{$\widetilde G=5$}\\ \hline
$\bar N$ & $\widetilde{E}_{\bar N}^{(e)}$  &  $\widetilde{E}_{\bar N}^{(app)}$
& $\widetilde{E}_{\bar N}^{(e)}$  &  $\widetilde{E}_{\bar N}^{(app)}$    
& $\widetilde{E}_{\bar N}^{(e)}$  &  $\widetilde{E}_{\bar N}^{(app)}$    &
$\widetilde{E}_{\bar N}^{(e)}$  &  $\widetilde{E}_{\bar N}^{(app)}$    \\
\hline
0           &  4.2872 &  4.2812       
            &  3.4245 &  3.4266
            &  3.1583 &  3.1601 
            &  3.1493 &  3.1510 \\ 
1           &  6.0892 &  6.1056    
            &  5.6302 &  5.6237
            &  5.3673 &  5.3621 
            &  5.3524 &  5.3472 \\
2           &  8.1171 &  8.1021     
            &  7.8422 &  7.8485
            &  7.6136 &  7.6190 
            &  7.5965 &  7.6015\\
3           &  10.266 &  10.271     
            &  10.103 &  10.101
            &  9.9314 &  9.9297 
            &  9.9157 &  9.9140 \\
\end{tabular}
\caption{Comparison between exact $(e)$ ``trapped'' solutions of 
Eq.(\ref{eqho}) and approximate $(app)$ ones, obtained from the ansatz
(\ref{parab}) for ${\cal N}=5$ levels. 
The coefficients $(a,b,c)$ of the parabola are (0.086,1.738,4.281) 
when $\widetilde{G}=0.05$, 
(0.014,2.183,3.427) when $\widetilde{G}=0.1$,
(0.027,2.175,3.160) when $\widetilde{G}=1$ 
and (0.029,2.167,3.151) when $\widetilde{G}=5$.}  
\end{center}
\end{table}

With the collective energy fixed, the coefficients $a$, $b$ and $c$ 
can be found. 
 
It should be reminded that, being the system 
of the equations (\ref{Vieta1}), (\ref{Vieta2}) and (\ref{Vieta3}) non-linear, 
more than one set of solutions is generally found: 
however the appropriate set is easily selected, being 
the one yielding energies in between the single particle levels. 

We quote in Table~3 our predictions for the eigenvalues of the pairing 
hamiltonian for one pair in the ${\cal N}=5$ case, 
using as input (\ref{Ecappr}) when $\widetilde G=1$ and $\widetilde G=5$ 
and (\ref{Ecubic}) when $\widetilde G=0.05$ and $\widetilde G=0.1$.
These leading orders are iterated via (\ref{Erecursion}) until 
self-consistency is reached. Our results are seen to agree with the exact ones 
obtained via the numerical solution of (\ref{eqho}) to better than $0.27\%$.   

In this letter, to pave the way to the problem of any number of fermions 
pairs, we have solved, almost analytically, the pairing problem for one pair 
living in any number of harmonic oscillator levels. 
Crucial for this achievement has 
been the conjecture (\ref{parab}), possibly related to the specific degeneracy 
of the harmonic oscillator single particle levels.
If this is true formulas like (\ref{parab}) can as well hold valid for others 
one body potentials, providing their degeneracy is known.

\end{document}